\documentclass[aps,prl,twocolumn,groupedaddress,showpacs,footinbib]{revtex4}

\usepackage{amssymb}

\usepackage{graphicx}

\begin{document}

\title{Quantum divisibility test and its application in mesoscopic
physics}

\author{G.B.\ Lesovik$^{a,b}$, M.V.\ Suslov$^{c}$, and G.\ Blatter$^{b}$}

\affiliation{$^{a}$L.D.\ Landau Institute for Theoretical Physics RAS,
   117940 Moscow, Russia}

\affiliation{$^{b}$Theoretische Physik, ETH-H\"onggerberg, CH-8093
   Z\"urich, Switzerland}

\affiliation{$^{c}$Moscow Institute of Physics and Technology,
   Institutskii per.\ 9, 141700 Dolgoprudny, Moscow District, Russia}
\date{\today}

\begin{abstract}
  We present a quantum algorithm to transform the cardinality of a set of
  charged particles flowing along a quantum wire into a binary number.  The
  setup performing this task (for at most $N$ particles) involves $\sim \log_2
  N$ quantum bits serving as counters and a sequential read out.  Applications
  include a divisibility check to experimentally test the size of a finite
  train of particles in a quantum wire with a one-shot measurement and a
  scheme allowing to entangle multi-particle wave functions and generating
  Bell states, Greenberger-Horne-Zeilinger states, or Dicke states in a
  Mach-Zehnder interferometer.
\end{abstract}


\maketitle

\section{Introduction}

Quantum mechanics offers novel algorithms allowing to speed up the solution of
specific computational tasks, some modestly, such as sorting a list
\cite{NielsonChuang,Grover}, while others, such as prime factorization
\cite{NielsonChuang,Shor}, are accelerated exponentially. While applications
in quantum cryptography \cite{NielsonChuang,BB} are close to commerical
realization \cite{Gisin}, the endeavour of building a universal quantum
computer with thousands of quantum bits lies in the distant future, if ever
realized. In this situation, it is interesting to consider special tasks which
are less demanding in their requirement with regard to the number of qubits
and the complexity of its network. An example of such an application is the
use of a qubit as a measuring device in the realization of full counting
statistics \cite{Hassler}; in a similar spirit, an iterative phase estimation
algorithm has been proposed as a testbed application for a limited amount of
qubits, in particular, a two-qubit benchmark \cite{Wendin}.  Here, we discuss
other applications where a few qubits serve as active or passive detectors.
The core element on the algorithmic side is a specific physical setup with $K$
qubits allowing to perform a (non-demolition) count of the elements $n <
N=2^K$ in a stream of particles flowing in a quantum wire, i.e., determining
its cardinality.  This algorithm resembles the phase estimation problem
\cite{Kitaev,Ekert} in inverted form: Rather then determining a phase $\phi$
with $N$ gate operations, here, the phase $\phi$ is known and we seek to find
the number $N$ of operations associated with the passage of the particles.
Our measurement scheme, involves a sequential readout, where the $j$-th
reading depends on the results of the previous $j-1$ measurements, reminding
about binary search trees \cite{Andersson}. A simpler, simultaneous (rather
then conditional) readout of the $K$ qubits provides a divisibility check (by
$2^K$) of the cardinality.  Combining the counter with a Mach-Zehnder
interferometer in a `which path' setup \cite{Heiblum}, we study interference
effects in the particle flow across the device and show how to make use of the
counter in the fabrication of entangled many-particle wave functions of
various kinds. With this program, we position ourselves at the interface
between information theory and its application in mesoscopic physics; rather
than universal, our quantum counting scheme is a special purpose algorithm,
admitting a relatively simple implementation while offering practical
applications.

The use of two-level systems as clocks or counters has a long history: using
the Larmor precession of a spin as a clock attached to the particle itself,
Baz' \cite{Baz} and Rybachenko \cite{Rybachenko} proposed to measure the time
it takes a particle to traverse a barrier in a tunneling problem. In the
context of full counting statistics in mesoscopic physics, Levitov and Lesovik
\cite{LevitovLesovik} introduced the idea to use an independent stationary
spin as a measurement device to count the electrons flowing in a nearby
quantum wire.  In quantum optics, Brune {\it et al.} \cite{Brune} proposed to
make use of atoms excited to Rydberg states as atomic clocks to count photons
in a cavity, a proposal that has been experimentally realized recently
\cite{Guerlin}; in this case, the flying atoms measure the number of localized
photons in the cavity. Our algorithm can be used to count photons as well; in
our dual setup the counters are fixed and (microwave) photon pulses propagate
in a transmission line.

\section{Counting Algorithms}

\subsection{Classical algorithm}

We start out with the counting problem, the transformation of the magnitude of
a set of (charged) particles (i.e., its cardinality $n$) into a binary number.
First consider the obvious classical algorithm and assume that each particle
passing a classical counter generates a pulse; to simplify the discussion, we
can assume taking the particle `$\bullet$' itself from the string. Consider a
register with $K$ ($n < 2^K$) empty slots $[0,0,\dots,0,0,0]$, then the first
particle is placed in the right most position $[0,0,\dots,0,0,\bullet]$, the
second induces a shift of the first to the next register
($[0,0,\dots,0,0,\bullet\bullet] \to [0,0,\dots,0,\bullet,0]$), the third
particle fills the first slot again ($[0,0,\dots,0,0,\bullet,\bullet]$), the
fourth particle induces two shifts ($[0,0,\dots,0,0,\bullet,\bullet\bullet]
\to [0,0,\dots,0,\bullet\bullet,0] \to [0,0,\dots,0,\bullet,0,0]$), etc.. The
transformation of the set's magnitude $n < N=2^K$ to a binary number then
involves $\sim n \log_2 n$ steps.

\subsection{Quantum measurement}

The simplest scheme using spin counters to determine the number of particles
flowing across a wire \cite{LevitovLesovik} requires $\propto n^2$
measurements and thus is even more demanding: to fix ideas, we assume
transport of charged particles along $x$ and $N_m$ spins initially polarized
along the positive $y$-axis. Upon passage of a charge, the induced $B$-field,
locally directed along the $z$-axis, rotates the spins in the $x$-$y$ plane by
a fixed angle $\phi < \pi/N$. In a real experiment, the spins could be
replaced by suitable qubits \cite{Hassler} and below, we will use the terms
`spin' and `qubit' synonymously.  The use of $N_m$ spins is equivalent to an
$N_m$-fold repetition of the same experiment with one spin, allowing us to
transfer the $n < N$ particles once and perform $N_m$ measurements on equally
prepared spins---this procedure corresponds to a single shot measurement of
$N_m$ spins (note, that the no-cloning theorem \cite{Wotters,Dieks} prevents
us from using one spin and then clone it after the passage of the $n$
particles). Measuring the spin along the $y$-axis, the (theoretical)
probability to find it pointing upwards is given by $P^\uparrow = \langle
m^\uparrow \rangle_k/N_m = \cos^2 (n\phi/2)$, where $\langle m^\uparrow
\rangle_k$ denotes the average of finding $m^\uparrow$ of the $N_m$ spins
pointing up in a sequence of $k \to \infty$ realizations of the entire
experiment. On the other hand, the one-time measurement $m_m^\uparrow$
provides the experimental result $P_m^\uparrow = m_m^\uparrow/N_m$, from which
we can find the number $n = (2/\phi) \arccos[(P_m^\uparrow)^{1/2}]$. A variant
of this scheme is proposed in Ref.\ \cite{Brune}, where a sequence of Rydberg
atoms brought into a quantum superposition through the interaction with cavity
photons is used to project the cavity-field onto a photon number state.

The above procedure is a statistical one and we have to determine how many
spins (measurements) $N_m$ are needed to predict the particle number $n$ with
certainty. The difference (we assume $N > n \gg 1$) $\delta P^\uparrow =
|P^\uparrow(n+1)-P^\uparrow(n)| \approx |\partial_n P^\uparrow| = (\phi/2)
\sin(n\phi)$ has to be much larger than the uncertainty $[\langle (\delta
m^\uparrow)^2\rangle_k]^{1/2} \equiv [\langle (m^\uparrow-\langle
m^\uparrow\rangle_k)^2\rangle_k]^{1/2}$ in the measurement, $\delta P^\uparrow
\gg [\langle(\delta m^\uparrow)^2\rangle_k]^{1/2}/N_m$.  Given the binomial
statistics of the measurement process (the values $\uparrow$ and $\downarrow$
are measured with probabilities $P^\uparrow$ and $(1-P^\uparrow)$), we obtain
$\langle (\delta m^\uparrow)^2\rangle_k = P^\uparrow (1-P^\uparrow) \, N_m$
and combining these results, we find that $N_m \gg 1/\phi^2 > N^2/\pi^2 \gg 1$
spins are needed in order to accurately measure the particle number $n < N$.
\begin{figure}[ht]
  \includegraphics[width=8.0cm]{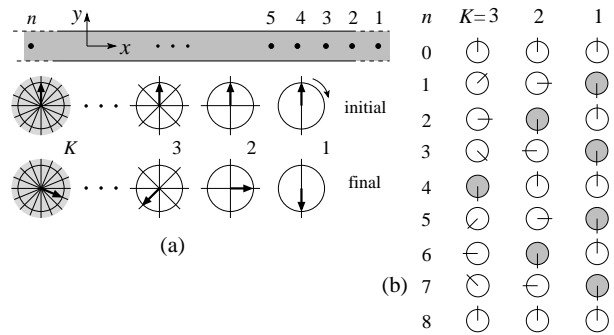}
  \caption[]{Illustration of the quantum algorithm to transform the
  cardinality $n$ of a set of particles into a binary number. (a) Initially,
  all spins point into the $+y$ direction. The $j$-th spin is rotated
  (clockwise) by $\phi_j = 2\pi/2^j$ upon passage of one particle. After the
  passage of all particles, the first spin is measured along the $y$-axis and
  provides the number's parity. Depending on the parity, the second spin is
  measured along the $y$-axis (even parity) or $x$-axis (odd parity); a
  measurement along (opposite to) the axis is encoded with a 0 (1). The
  further iteration is straightforward: depending on the previous outcomes of
  the $(j-1)$ measurements, the $j$-th spin is measured along one of the
  angles $l\phi_j$ with $l \in \{0,\dots,j-1\}$ and the $j$-th position in the
  binary register assumes values 0 or 1 depending on the measurement result.
  The figure shows the reading after passage of 5 particles with $K=4$.  (b)
  Divisibility check: qubit states after passage of $n=0,\dots, 8$ electrons
  for $K=3$; for $n=1,\dots, 7$ there is exactly one qubit ending up in the
  $|\!  \downarrow\rangle$-state (shaded), signalling that the cardinality $n$
  of the sequence is not divisible by $2^3=8$.}
  \label{fig:counting}
\end{figure}

\subsection{Quantum algorithm}

We now consider a more sophisticated measurement scheme, assuming the role of
a quantum algorithm, where we need only a number $K \sim \log_2 N$ of spins to
encode the magnitude of a set with $n < N$ particles into a binary number. The
$K$ spins are all pointing up initially, see Fig.\ \ref{fig:counting}. Upon
passage of a particle, the $j$-th spin is rotated (clockwise) by the amount
$\phi_j = 2\pi/2^j$ (rotation by $U_z (-\phi_j)$); such different rotation
angles are implemented through different coupling strengths of the
spins/qubits to the wire, cf.\ Ref.\ \onlinecite{Hassler}. The passage of $n$
particles rotates the $j$-th spin by the amount $n \phi_j$. In particular, the
first spin is rotated by the angle $n\pi$ and either points upward if the
number's parity is even (we store a `0' in the first position of the binary
number) or downward (we store a `1' in the first position of the binary
number) if the parity is odd. Hence the measurement of the first spin along
the $y$-axis provides already the parity of the number (note that we
had to perform $n \log_2 n$ operations in the classical algorithm to find the
parity). In addition, the first spin will determine the axis in the
measurement of the second spin: for an even $n$, we measure $\phi_2$ along the
$y$-axis (and store a 0 if the spin is pointing up and a 1 if the spin is
pointing down), while for an odd-parity $n$, we measure $\phi_2$ along the
$x$-axis.  More specifically, if $n=4l_2$, with $l_2$ the number of full
rotations of the spin number 2, the state of the first spin signals even
parity and the second spin, measured along the $y$ direction, points into the
direction $+y$, hence we store a `0' in the second position of the binary
number.  Similarly, for $n = 4l_2+1$, the spin 2 is directed along $+x$; spin
1 signals odd parity, the measurement is done along the $x$-axis, and we store
a `0'.  For $n = 4l_2+2$, spin 2 is directed along $-y$ (even parity,
measurement along $y$, store `1'), and for $n = 4l_2+3$ the spin 2 is directed
along $-x$ (odd parity, measurement along $x$, store `1').

The iteration of the algoritm is straightforward: the $j$-th spin is measured
along the angle $m_{j-1} \phi_j$ with the integer $m_{j-1}$ corresponding to
the binary number encoded in the $j-1$ previous measurements.  The $j$-th
position in the binary register then assumes a value 0 or 1 depending on the
measurement result, 0 for a spin pointing along the axis and 1 for a spin
pointing opposite. The entire algorithm requires $\lceil \log_2 (n+1) \rceil$
steps ($i_r=\lceil r \rceil$ is the closest integer $i_r > r$), the same
number as bits required to store the number $n$ in binary form, and provides
an exponential speedup compared with the classical algorithm
\footnote{Strictly speaking, this statement is valid if the unit time step in
the algorithm is larger than the time separation between particles.}. 

\section{Single Shot Divisibility Check}

A variant of the above counting algorithm is a test for divisibility by powers
of two: given a finite train of electrons propagating in a wire, we wish to
check whether the number of electrons in the train (its cardinality) is
divisible by $2^K$. Obviously, the information on the divisibility of the
train's cardinality by $2^K$ is reduced as compared to the information on its
cardinality; correspondingly, we expect a reduced effort to achieve this task.
Indeed, using the above setup, the divisibility check involves $K$ qubits and
their simultaneous measurement along the $y$-axis at the end of the train's
passage (rather than the conditional measurement above). The train's
cardinality is divisible by $2^K$, if all spins are pointing up, i.e., along
the positive $y$-axis, cf.\ Fig.\ \ref{fig:counting}(b); the non-divisibility
is signalled by the `opposite' outcome, i.e., there is at least one spin
pointing down.

The above statement relies on the fact, that after the passage of $n=2^K$
particles all counters end up in the spin-up state, while for $n \neq 2^K$
there is exactly one spin residing in a spin-down state, cf.\ Fig.\
\ref{fig:counting}(b) (note the difference in having counters in up/down
states with well defined measurement outcomes and statistical results of
up/down measurements for counters pointing away from the $y$ direction).  We
provide a formal proof of this statement: starting in the initial state (with
the quantization axis along $z$) $|\mathrm{in}\rangle= |\!+y\rangle=
(|\!\uparrow\rangle + i|\!\downarrow\rangle)/\sqrt{2}$, after passage of $n$
particles, the $j$-th spin ends up in the final state $|\mathrm{f}\rangle =
[e^{\pi i n/2^j} |\!\uparrow\rangle + i e^{-\pi i n/2^j}
|\!\downarrow\rangle]/\sqrt{2}$. The probability to measure this spin along
the $+y$-direction is $|\langle\!+y |\mathrm{f}\rangle|^2 = \cos^2 (\pi
n/2^j)$, $j=1,\dots, K$. There is exactly one spin $1 \leq j^\ast \leq K$, for
which this probability vanishes: this follows from the statement, that any
number $0 < n < 2^K$ can be represented in the form $2^m I$ with $0 \leq m <
K$ and $I$ an odd integer. Then, for the spin $j^\ast = m+1$ (and only for
this spin) the phase $\pi n/2^{j^\ast} = \pi I/2$ is an odd multiple of
$\pi/2$ and hence the probability $|\cos(\pi I/2)|^2$ to find it pointing
along $+y$ vanishes, i.e., the spin is pointing down. For all other spins $j
\neq m+1$, the phase is a multiple of $\pi$ (for $j<m+1$, the spin is pointing
up) or a fraction $I/2^{j-m-1}$ of $\pi/2$ (for $j > m+1$, the spin is not
pointing down).  

\section{Implementation}

The spins required in the above counting- and divisibility-check algorithms
can be implemented using various types of qubits; note that, while the special
nature of our algorithm avoids the large number of qubits and huge network
complexity of a general-purpose quantum computer, we do require individual
qubits with high performance.  Most qubits naturally couple to the electrons
in the quantum wire, either via the gauge field (current) or via the scalar
potential (charge). The coupling to charge is strong, with typical rotation
angles $\phi$ of order $(e^2/\hbar v_{\rm \scriptscriptstyle F}) \ln(L/d)$,
with $L$ the wire's length, $d$ its distance from the qubit, and $v_{\rm
\scriptscriptstyle F}$ the Fermi velocity, thus allowing for a $\pi$-phase
rotation upon passage of one unit of charge.  Transverse coupling via the
current is weak, usually requiring enhancement with a flux transformer, cf.\
the discussion in Ref.\ \onlinecite{Hassler}. 

To fix ideas, below we discuss an implementation with charge qubits in the
form of double quantum dots (DQD) as one attractive possibility which can be
manipulated via electronic gates and offers various modes of operation.
DQDs have been implemented in GaAs/AlGaAs heterostructures
\cite{Hayashi,Petta} or as an isolated (leadless) version in Si technology
\cite{Gorman}, the former with typical oscillation frequencies in the few GHz
regime and nano-second phase decoherence times, resulting in quality factors
of order $1$ to $10$; characteristic tunneling couplings/decoherence times are
a factor 100 smaller/larger in the isolated qubit \cite{Gorman}.
Alternatively, one may consider superconducting charge qubits, e.g., the
`Quantronium' \cite{Vion}, with a decoherence time reaching nearly a $\mu$s;
this value, measured at the `sweet spot', will be reduced, however, when
choosing a working point which is suitably sensitive to charge. At present,
the resolution in the competition between a suitable charge sensitivity to
achieve rotation angles of order $\pi$ and the decoherence due to fluctuating
charges in the environment remains a technological challenge. On the other
hand, todays best solid state qubits (with a decoherence time above 2 $\mu$s),
the transmon \cite{Koch,Schreier}, could be used as photon counters in the
microwave regime \cite{Blais}.

The above qubit characteristics have to be compared with the typical time
scale of electronic transport in the wire. While under dc bias conditions,
subsequent electrons are separated by the voltage time $\tau = h/eV$,
single-electron wave packets can be generated by unit-flux voltage pulses of
Lorentzian shape \cite{LLL,Keeling}. Recently, an alternative scheme has been
used by Feve {\it et al.} \cite{Feve}, who have injecting individual electrons
from a quantum dot into an edge channel formed in the quantum Hall regime.
Typical time scales $\tau = h/T \delta$ of single-electron pulses in their
experiment range between 0.1 and 10 nano-seconds \cite{Feve}, where $T$ and
$\delta$ denote the tunneling probability and the level separation between
states in the dot feeding the quantum wire. We conclude that today's charge
qubits are at the border of becoming useful for the proposed electron counting
experiments.
\begin{figure}[ht]
  \includegraphics[width=7.5cm]{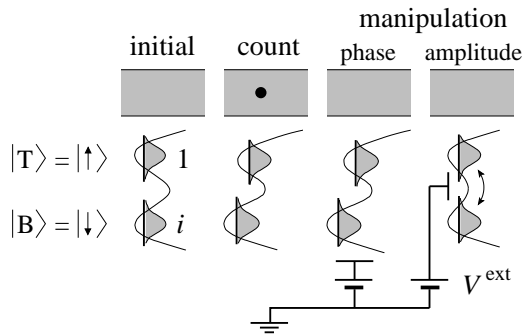}
  \caption[]{Implementation of the counting algorithm with double-dot charge
  qubits modelled as double-well systems. Initial state $|+y\rangle =
  [{|\!\uparrow\rangle} +i{|\!\downarrow\rangle}]/\sqrt{2}$. Particles are
  counted via their associated voltage pulses generating a phase shift between
  the states $|\mathrm{T} \rangle$ and $|\mathrm{B}\rangle$ (rotation around
  $z$).  The initalization and readout involve manipulations of phase
  (disbalancing the levels $|\mathrm{T} \rangle$ and $|\mathrm{B}\rangle$,
  rotation around $z$) and of amplitude (lowering the barrier between
  $|\mathrm{T} \rangle$ and $|\mathrm{B}\rangle$, rotation around $x$).}
  \label{fig:dd}
\end{figure}

We assume the two dots aligned perpendicular to the wire, such that they
couple differently to the electron charge in the wire and model the double dot
as a two-well potential with quasi-classical states $|\mathrm{T}\rangle \equiv
|\!\uparrow\rangle$ (top well, see Fig.\ \ref{fig:dd}; we use spin language in
our analysis below) and $|\mathrm{B}\rangle  \equiv |\!\downarrow\rangle$
(bottom well) and ground/excited states $|\pm\rangle = [{|\!\uparrow\rangle}
\pm {|\!\downarrow\rangle}]/\sqrt{2}$ separated by the gap $\Delta$. We first
consider a `phase mode operation' of the counter.  Assuming a large barrier
separating the quasi-classical states, the tunneling amplitude $\propto
\Delta$ is exponentially small.  In order to prepare the qubits in the state
$|+y\rangle = [{|\!\uparrow\rangle} +i{|\!\downarrow\rangle}]/\sqrt{2}$, we
measure their states and subsequently rotate them around $x$ by an angle
$-\pi/2$ ($\pi/2$) if the state $|\!\uparrow\rangle$ ($ |\!\downarrow\rangle$)
was measured.  The rotation around $x$ involves a lowering of the barrier
separating the quasi-classical states, allowing for an amplitude shift between
them, cf.\ Fig.\ \ref{fig:dd}: the opening of a finite gap $\Delta$ during the
time $t$ adds an additional phase evolution $e^{-i\Delta t/\hbar}$ to the
excited state $|-\rangle$ and thus corresponds to a rotation of the spin
around the $x$-axis by the angle $\Delta t/\hbar$; choosing a time $t = \hbar
\pi/2\Delta$ generates a rotation of the state $|\!\downarrow\rangle$ to the
state $|+y\rangle$.  Alternatively, the qubits are relaxed to the ground state
$|+\rangle$ (corresponding to a spin pointing along $+x$) and subsequently
rotated by $\pi/2$ around the $z$-axis via a suitable bias pulse applied to
the double-dot, adding the relative phase $\pi/2$ to the quasi-classical state
$|\!\downarrow\rangle$, cf.\ Fig.\ \ref{fig:dd}.

The passage of electrons in the wire generates a final state $|\Psi\rangle =
[|\!\uparrow\rangle +i e^{-i\phi_j n}|\!\downarrow\rangle]/\sqrt{2}$, where
$\phi_j=2\pi/2^j$ is the properly tuned phase difference picked up by the
quasi-classical states upon passage of one electron. The readout step for the
divisibility check involves a rotation around the $x$-axis by an angle
$\pi/2$.  The divisibility check then tests for the presence of all dot
electrons in the state $|\!\uparrow\rangle$; if the answer is positive, the
number $n$ of particles passing the $K$ double-dots is divisible by $2^K$. In
order to find the exact value of the cardinality $n$, another rotation around
the $z$-axis by an angle $m_{j-1}\phi_j$ has to be performed before rotating
around $x$, where the integer $m_{j-1}$ corresponds to the binary number
encoded in the first $j-1$ measurement outcomes; e.g., for the third qubit
$j=3$, after passage of 7 electrons, the measurement of the first two qubits
provides the binary number $(1,1)$, hence $m_2 = 3$, and a rotation by
$3\pi/4$ around $z$ makes the third spin point along the $-y$ direction;
storing a 1 as the third digit of the binary number we obtain $m_3 =7$, cf.\
Fig.\ \ref{fig:counting}(b).

Using the above phase mode operation, the double dot does not act back on the
passing electrons in the wire, since the charge distribution remains unchanged
during all of the detection phase. Another version of the divisibility check
makes use of the back action of the double dot on the wire and tests for the
divisibility without explicit measurement of any of the final qubit states.
This gain in performance has to be traded against two disadvantages: first,
the backaction has to be properly controlled, and second, the particle train
has to be properly sequenced in time, following a prescribed time separation
between two consecutive particles.

\begin{figure}[ht]
  \includegraphics[width=8.0cm]{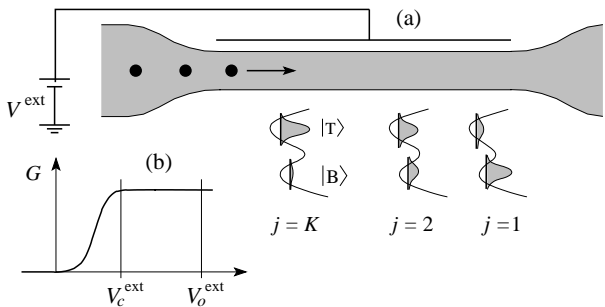}
  \caption[]{(a) Setup for divisibility check with `self-measurement'. The
  disbalanced states $|\mathrm{T}\rangle$ and $|\mathrm{B}\rangle$ of the
  two-level charge qubits are brought into degeneracy when electrons pass the
  QPC and the time evolution moves a fraction of the qubit's wave function
  between $|\mathrm{T}\rangle$ and $|\mathrm{B}\rangle$. In turn, the qubits'
  charges act back on the QPC and modify its conductance $G$ via narrowing
  (wave function at the top) and widening (wave function in the bottom) the
  constriction.  (b) External bias set to $V^{\mathrm{ext}}_c$ with the
  channel barely open und to $V^{\mathrm{ext}}_o$ defining a wide open
  channel.}
  \label{fig:div_check}
\end{figure}

The setup then involves a quantum point contact (QPC) that can be manipulated
through an external gate $V^\mathrm{ext}$, see Fig.\ \ref{fig:div_check}(a).
This time, the double dot qubits are prepared with asymmetric states (with
energy difference $\varepsilon$) in the unbiased situation. Initially, each
qubit is in the (high energy) $|\mathrm{B}\rangle$ state with the electron
further away from the wire.  The passage of an electron in the wire brings the
two states $|\mathrm{T}\rangle$ and $|\mathrm{B}\rangle$ into degeneracy and a
fraction of the wave function tunnels from $|\mathrm{B}\rangle$ to
$|\mathrm{T}\rangle$. The role of the angle $\phi$ is now played by the phase
$\phi \approx \Delta \delta t_\mathrm{deg}/\hbar$, with $\Delta$ the tunneling
gap and $\delta t_\mathrm{deg}$ the degeneracy time (in reality, the time
evolution of the electric potential due to the passing electron has to be
properly accounted for). In order to assure proper evolution of the DQD's
wavefunction due to the passage of subsequent electrons, the trivial phase
evolution in between $|\mathrm{T}\rangle \leftrightarrow |\mathrm{B}\rangle$
tunneling events has to be an integer ($k$) multiple of $2 \pi$, $\varepsilon
t_\mathrm{con} /\hbar \approx 2k\pi$.

The electrons in the qubits act back on the quantum wire through a capacitive
coupling and can block the channel. We define the `critical'
($V^\mathrm{ext}_c$) and the `open' ($V^\mathrm{ext}_o$) bias settings of the
external gate in the following way (see Fig.\ \ref{fig:div_check}(b)): With
all qubits in the $|\mathrm{B}\rangle$ state, we tune the QPC to one
transmitting channel barely open such that the shift of the electron in one of
the qubits to the $|\mathrm{T} \rangle$ state suffices to block the
channel---this defines $V^\mathrm{ext}_c$.  On the other hand, setting the
bias to $V^\mathrm{ext}_o$ widely opens the transmitting channel such that the
electrons move with appreciable velocity through the channel. Given these two
settings, the divisibility check is easily implemented: We apply a bias
$V^\mathrm{ext}_o$ to the external gate and let the particle train pass the
QPC.  Subsequently, we switch the external gate to its critical value
$V^\mathrm{ext}_c$ and send one more (test) electron through the QPC.  If the
cardinality of the train is divisible by $2^K$, then all qubits have returned
back to the $|\mathrm{bottom} \rangle$ state, the (test) electron can pass the
QPC and is detected on the other side, e.g., via a single electron transistor.
On the other hand, if the cardinality of the train is not divisible by $2^K$,
then exactly one of the $K$ qubits is in the $\mathrm{top}$ state, see Fig.\
\ref{fig:div_check}(c), and definitively blocks the channel (while other
qubits may reflect particles only indeterministically).  This scheme provides
a single shot test for the divisibility by $2^K$ of the particle train's
cardinality.

Note that the channel has to be wide open during the transmission of the
electrons in order to prevent their entanglement with the counter-qubits
through back action: the scalar interaction ${\cal V}$ between the qubit and
the electron in the wire decelerates the latter. This deceleration generates a
time delay $t_\mathrm{del} = \int dx \{ 1/v[{\cal V}(x)] - 1/v[{\cal V}=0] \}$
which depends on the charge state of the qubit, hence the two qubit states
$|\mathrm{T} \rangle$ and $|\mathrm{B} \rangle$ are entangled with
portions of the particle's wave packet which are delayed in time.  We then
require the electron to move fast through the channel and thus demand that the
QPC be biased away from criticality, implying a weak backaction and hence
a negligible time delay.


\subsection{Self-organized bunching}

\begin{figure}[ht]
  \includegraphics[width=7.0cm]{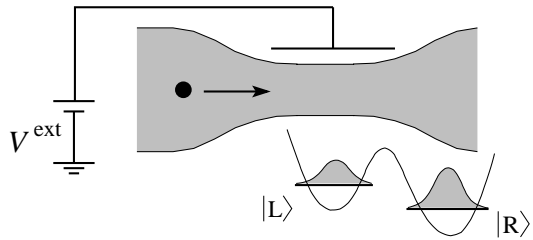}
  \caption[]{Setup for bunching. The passage of an electron in the wire brings
  the states $|\mathrm{L}\rangle$ and $|\mathrm{R}\rangle$ into
  degeneracy and the time evolution shifts a fraction $\phi/\pi$ of the wave
  function to the other level.  The qubit's charge acts back on the QPC and
  changes its conductance $G$, blocking the channel when residing in the
  $|\mathrm{L}\rangle$ state and opening the channel after tunneling to the
  $|\mathrm{R}\rangle$ state. Note that in the present geometry, reflected
  particles do not bring the qubit into resonance and hence do not modify the
  relative weights in the qubit's wave function.}
  \label{fig:bunching}
\end{figure}

A modified setup of the qubit-controlled quantum point contact can be used to
generate self-organized bunching, however, this setup involves strong
backaction and is difficult to control (see Ref.\ \onlinecite{Snyman} for a
recent study where a strongly coupled qubit modifies the transport through a
QPC). The basic idea is, that while a simple quantum scatterer (a tunneling
barrier) transforms a regular stream of particles into a perfectly random
sequence with binomial statistics (transmission versus reflection), our
qubit-controlled QPC will generate a non-trivial, tunable, and non-Markovian
random sequence.  We consider the simplest case with one qubit controlling the
QPC, cf.\ Fig.\ \ref{fig:bunching}, and start with an initial state where the
qubit electron resides in the high-energy left state $|\mathrm{L}\rangle$ and
the quantum point contact is barely closed through a critical tuning of the
QPC with the external gate $V^\mathrm{ext}$.  The passage of one electron
brings the right state $|\mathrm{R}\rangle$ into resonance with
$|\mathrm{L}\rangle$ and the qubit electron tunnels to the state
$|\mathrm{R}\rangle$ away from the QPC (we assume a phase angle $\phi = \pi$).
The QPC then is open and as the next electron flows down the channel, the
qubit electron tunnels back to $|\mathrm{L} \rangle$, thus closing the channel
again (we assume a properly time sequenced flow).  Adding a second control
qubit (e.g., on the other side of the QPC, with the channel blocked when both
qubit electrons reside in the $|\mathrm{L}\rangle$ state) with phase angle
$\phi=\pi/2$, bunched electron trains with four particles can be formed.  The
train can be initiated randomly through tunneling of the initial electron or
in a controlled way via a voltage pulse.
The complexity of the system's evolution is already appreciated for the case
with only one qubit controlling the QPC (we denote the left (right) state by
$|\!\uparrow\rangle$ ($|\!\downarrow\rangle$)). The incoming electron $|{\rm
in}\rangle= |\Psi_0\rangle$ is transmitted across ($|t\rangle$) or reflected
by ($|r\rangle$) the QPC, depending on the state of the qubit,
\begin{eqnarray} \label{eq:ab}
  &&|{\rm in}\rangle |\sigma\rangle \to |t\rangle
       \sum_{\sigma^\prime} t_{\sigma\sigma^\prime} |\sigma^\prime\rangle
                                       +|r\rangle
       \sum_{\sigma^\prime} r_{\sigma\sigma^\prime} |\sigma^\prime\rangle.
\end{eqnarray}
Hence, after scattering of one electron, the system's initial state $|{\rm
in}\rangle (a_i |\!\uparrow\rangle + b_i |\!\downarrow\rangle)$ evolves to the
final state $a_f |t\rangle |\!\uparrow\rangle +b_f|t\rangle |\!\downarrow\rangle
+c_f|r\rangle |\!\uparrow\rangle +d_f|r\rangle |\!\downarrow\rangle$ with
\begin{eqnarray} \nonumber
  a_f = a_i t_{\uparrow\uparrow} + b_i t_{\downarrow\uparrow}, &\quad&
    b_f = a_i t_{\uparrow\downarrow} + b_i t_{\downarrow\downarrow},
  \\ \nonumber
    c_f = a_i r_{\uparrow\uparrow} + b_i r_{\downarrow\uparrow}, &\quad&
    d_f = a_i r_{\uparrow\downarrow} + b_i r_{\downarrow\downarrow}.
\end{eqnarray}
For an ideal setup, we have $|r_{\uparrow\uparrow}| = |t_{\downarrow\uparrow}|
= 1$ and all other coefficients vanish, hence the final state assumes the
simple form $a_i r_{\uparrow\uparrow} |r\rangle |\!\uparrow\rangle + b_i
t_{\downarrow\uparrow} |t\rangle |\!\uparrow\rangle$. Further extensions
beyond two qubits are more difficult to realize, as the qubits have to act
jointly on the QPC; this will introduce uncontrolled interaction effects among
the qubits, perturbing their proper `rotation'.  Also, keeping the QPC close to
criticality, the velocity of the transmitted electrons depends on the qubits'
states, which thus get entangled with correspondingly time-delayed portions of
the wave function. The above idealized bunching will then give way to some
self-organized bunching (a non-Markovian process) which might be interesting
in itself, though not perfectly controlled.  

So far, we have discussed the setup of the `quantum cardinality-counter' and
its application to the manipulation of classical information; below we use
these ideas to control and modify quantum information.

\section{Manipulation of wave functions} 

\begin{figure}[ht]
  \includegraphics[width=7.0cm]{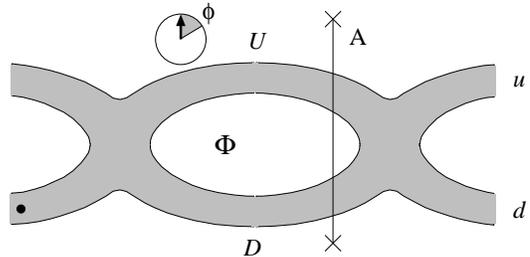}
  \caption[]{Mach-Zehnder interferometer with spin counter. Particles enter
  the interferometer through the left leads (here the bottom lead) and
  are measured on the right. The spin counter in the upper arm $U$
  detects the passage of particles via a rotation by the angle $\phi$. The
  magnetic flux $\Phi$ through the loop allows to tune the phase difference
  when propagating along different arms.}
  \label{fig:interferometer}
\end{figure}

Next, we discuss the manipulation of one- and two-body wavefunctions by a
spin/qubit counter.  Consider a particle entering the Mach-Zehnder
interferometer, see Fig.\ \ref{fig:interferometer}, from the lower-left lead
and exiting the loop through the upper-right lead $u$ where it is measured.
We denote the initial state injected into the interferometer by
$|\mathrm{in}\rangle = |\psi_0\rangle$.  The wave function can propagate along
two trajectories, the upper arm $U$ where the particle picks up a phase
$\varphi_U$ and the spin counter is flipped (we choose a rotation angle
$\phi=\pi$), or the lower arm $D$ accumulating a phase $\varphi_D$ and leaving
the spin unchanged.  Assuming symmetric splitters with transmission $t$
($|t|^2 = 1/2$) and reflection $r=\pm it$, the projection of the wave function
in the upper outgoing channel $u$ reads
\begin{eqnarray}
   \label{eq:Psi1}
   \Psi_u &=& t\,r\, e^{i\varphi_U} |\mathrm{in}, \downarrow\rangle 
            + r\,t\, e^{i\varphi_D} |\mathrm{in},\uparrow\rangle
   \\ \nonumber
   &=& (\pm i)\bigl\{e^{i\varphi_U}|\mathrm{in},\downarrow\rangle 
              + e^{i\varphi_D}|\mathrm{in},\uparrow\rangle\bigr\}/2.
\end{eqnarray}
A partial summation over the spin states provides us with the particle's
density matrix $\rho_u  = |\mathrm{in}\rangle \langle \mathrm{in}|/2$.  The
result tells us that all interference effects are gone due to the decoherence
by the spin counter: the visibility $V \equiv [\max P_u -\min P_u] /[\max
P_u+\min P_u]=\cos(\phi/2)$ of the oscillations in the probability $P_u =
|\Psi_u|^2$ vanishes, while the spin carries the full information ($I$) on the
particle's path \cite{Englert}.  Choosing another rotation angle $\phi$ for
the spin counter, the visibility can be tuned to any value between zero and
unity, with the conjugate behavior of the information gain by the counter,
$V^2(\phi)+I^2(\phi)= 1$, cf.\ Ref.\ \onlinecite{Heiblum}, where a similar
behavior has been observed in a `which path' experiment.

Next, we send two particles into the Mach-Zehnder (MZ) loop with the
spin-counter flipping by $\phi=\pi$ upon passage of one particle in the upper
arm.  We assume the two wave functions describing the initial state
$|\mathrm{in}\rangle = |\psi_{01} \psi_{02}\rangle$ to be well separated in
space, allowing us to ignore exchange effects in our (MZ) geometry. The part
of the wave function with two particles measured in the top-right arm then
reads
\begin{eqnarray}
   \label{eq:Psi2}
   \Psi_{2u} 
      &=& [-1] t^2r^2 e^{2i\varphi_U} |\mathrm{in},\uparrow\rangle
        +t^2r^2\, e^{i\varphi_U} e^{i\varphi_D} |\mathrm{in}, \downarrow\rangle
        \\ \nonumber
        &&\qquad
        +r^2t^2 \,e^{i\varphi_D} e^{i\varphi_U} |\mathrm{in}, \downarrow\rangle
        +r^2t^2 e^{2i\varphi_D} |\mathrm{in}, \uparrow\rangle
         \\ \nonumber
      &=& (-1)\bigl\{[-1]e^{2i\varphi_U}|\mathrm{in}, \uparrow\rangle 
        + e^{2i\varphi_D} |\mathrm{in}, \uparrow\rangle
        \\ \nonumber
      &&\qquad
        + 2 e^{i(\varphi_U+\varphi_D)}|\mathrm{in},\downarrow\rangle\bigr\}/4.
\end{eqnarray}
The factor $[-1]$ accounts for the phase $\pi$ picked up in the rotation of
the spin-state by $2\pi$ (for a qubit, this phase can be tuned and assumes the
value $[-1]$ if the coupling shifts the qubit levels symmetrically up and
down), while the factor $(-1)$ accounts for the additional scattering phases
$(\pm i)$ in the reflection process.  This time, the interference partly
survives; the summation over the spin states provides us with the probability
$P_{2u}$ to detect both particles in the upper arm,
\begin{eqnarray}
   \nonumber
   P_{2u} &=& P_{2u\uparrow}+P_{2u\downarrow} 
          = |[-1]e^{2i\varphi_U}+e^{2i\varphi_D}|^2 
          \langle\uparrow\!|\!\uparrow\rangle/16
          \\ \nonumber
          &&\qquad\qquad\qquad\quad
          + |2 e^{i(\varphi_U+\varphi_D)}|^2
          \langle\downarrow \!| \!\downarrow \rangle/16
          \\ \label{eq:rho2}
          &=& (1+[-1]\cos[2(\varphi_U-\varphi_D)])/8 + 1/4.
\end{eqnarray}
The two interference terms behave quite differently: in the first one
(associated with the unrotated spin $|\!\uparrow\rangle$), the phases
accumulated by the two particles when both travel along the upper/lower arms
add up and we observe a two-particle interference pattern (the occurrance of
two-particle interference in a Hanbury-Brown Twiss interferometer has been
proposed by Yurke and Stoler \cite{Yurke} and by Samuelsson {\it et al.}
\cite{SamBut_03} and observed in an experiment by Neder {\it et al.}
\cite{Heiblum_07}). The second term (associated with the flipped spin
$|\!\downarrow\rangle$) describes particles travelling in different arms and
the phases picked up along the upper and lower arms cancel.  Nonetheless, we
find that we have constructive interference with a doubled total probability
$1/4$ (the maximal value of the first term), but no Aharonov-Bohm oscillations
show up.

When calculating the average number of particles detected in the upper lead,
we have to add the probability resulting from those trajectories where only
one particle leaves the device through the lead $u$. Keeping track of the
out-terminals with the index $u$ or $d$, we obtain the corresponding part of
the wave function
\begin{eqnarray}
   \nonumber
   \Psi_{1u} &=&  r\,t \bigl\{[-1] 
      t^2 e^{2i\varphi_U} \bigl[|\mathrm{in},u,d,\uparrow\rangle
    + |\mathrm{in},d,u,\uparrow\rangle\bigr]
          \\ \nonumber && 
    + (r^2+t^2)\, e^{i(\varphi_U+\varphi_D)} 
            \bigl[|\mathrm{in},u,d,\downarrow\rangle
    + |\mathrm{in},d,u,\downarrow\rangle\bigr]
          \\ \nonumber && 
    + r^2 e^{2i\varphi_D} \bigl[|\mathrm{in},u,d,\uparrow\rangle 
    + |\mathrm{in},d,u,\uparrow\rangle\bigr] \bigr\}.
          \label{eq:Psi11}
\end{eqnarray}
Extracting the component associated with the up state of the spin-counter, we
find the probability
\begin{eqnarray}
   \label{eq:rho11}
   P_{1u\uparrow} &=& 2 \big|[-1]e^{2i\varphi_U}+(-1)e^{2i\varphi_D}\big|^2/16
          \\ \nonumber
          &=& (1+[-1](-1)\cos[2(\varphi_U-\varphi_D)])/4.
\end{eqnarray}
The total particle number $N_u$ measured in the upper lead and associated with
the spin-up state is given by $N_u = 2\, P_{2u\uparrow} + 1\, P_{1u\uparrow}
=1/2$ and the interference term cancels out.

\section{Projective multi-qubit entanglement}

The standard way to entangle quantum degrees of freedom makes use of
interaction between the constituents. An alternative is provided by a
projection technique, where a measurement selects the desired entangled state.
In some cases, the projection makes use of the entangled state but
simultaneously implies its destruction---more useful for quantum information
processing are those schemes which entangle qubits for further use after the
projection.  Examples for the latter have been proposed using various
arrangements of double-dot charge qubits combined with a quantum point contact
serving as a quadratic detector \cite{RuskovKorotkov, Trauzettel} or (free)
flying spin-qubits tracked via a charge detector, where the charge provides an
additional non-entangled degree of freedom associated with the entangled spins
\cite{BoseHume,Beenakker}. Here, we generate multi-qubit orbital entanglement
of flying qubits via their entanglement with our spin counters serving as
ancillas; after reading of the counter states, the entangled multi-qubit state
can be further used.

The setup in figure \ref{fig:interferometer} conveniently
lends itself for the generation of entanglement. The simplest example is
provided by the two-particle propagation analyzed above: Evaluating the wave
function at the position A before mixing in the second splitter, we find the
expression (cf.\ Eq.\ (\ref{eq:Psi2}); we use scattering coefficients for a
symmetric beam splitter, e.g., $t^2 = 1/2$ and $r^2 = (-1)/2$ and $r\,t=(\pm i)/2$
for the two particles injected from the bottom left)
\begin{eqnarray}
   \nonumber
   &&\Psi_{2A}
      =
      \bigl\{\bigl[[-1]e^{2i\varphi_U}|\!\Uparrow, \Uparrow \rangle 
      + (-1)e^{2i\varphi_D} |\!\Downarrow,\Downarrow\rangle\bigr] 
       \otimes |\!\uparrow\rangle
          \\ \label{eq:Psi2A}
      &&\quad
      (\pm i)\, e^{i(\varphi_U+\varphi_D)}\bigl[|\!\Downarrow,\Uparrow \rangle
                                           +|\!\Uparrow,\Downarrow \rangle\bigr]
               \otimes |\!\downarrow\rangle \bigr\}/2,
\end{eqnarray}
where we have introduced a pseudo-spin notation to describe the propagation of
the two particles along the two arms: a spin $\Uparrow$ ($\Downarrow$) refers
to propagation in the upper (lower) arm. Choosing $\varphi_U = \varphi_D$, the
measurement of the spin-counter in the $\uparrow$-state projects the particle
wave function to the Bell state $|\!\Uparrow,\Uparrow \rangle+ |\!\Downarrow,
\Downarrow\rangle$, while the measurement of the $\downarrow$-state generates
the state $|\!\Downarrow,\Uparrow \rangle +|\!\Uparrow,\Downarrow \rangle$.
The remaining two Bell states can be obtained by injecting the two particles
through the different leads on the left: The state $|\!\Uparrow, \Uparrow
\rangle$ ($|\!\Downarrow,\Downarrow \rangle$) then involves the coefficient
$r\,t$ ($t\,r$) rather than $t^2$ ($r^2$) and hence we obtain a minus sign in
the combination $|\!\!\Uparrow,\Uparrow \rangle -|\!\!\Downarrow, \Downarrow
\rangle$ (and similar for {$|\!\Downarrow,\Uparrow \rangle
-|\!\Uparrow,\Downarrow \rangle$}, which now involves the coefficients $t^2$
and $r^2$ rather than the factor $r\,t$ before).  Alternatively, one may
thread a flux $\Phi$ through the loop in order to manipulate the relative
phase $\varphi_D - \varphi_U = 2\pi\Phi/\Phi_0$, with $\Phi_0 = hc/e$ the unit
flux, in the state $|\!\Uparrow, \Uparrow \rangle + \exp(2\pi i \Phi/\Phi_0)
|\!\Downarrow, \Downarrow\rangle$. Note that the indistinguishability of
particles exploited in the above entanglement process is an `artificial' one
defined by the qubit detector, rather than the `fundamental' one of identical
particles.

The above scheme for entangling two particles with one spin-counter is easily
extended to $2^K$ particles and an array of $K$ spin-counters measuring the
cardinality of the particle set flowing through the upper arm. As an
illustration we consider the case $K=2$, four particles and two spin-counters.
We use the shorthand $|j\rangle$, $j=0,1,2,3,4$ with the identification
$|0\rangle = |4\rangle = |\!\uparrow,\uparrow\rangle$ for the four different
counter states, assume again a symmetric splitter, $\varphi_U = \varphi_D$,
and injection from the bottom left; then
\begin{eqnarray}
   \label{eq:Psi4A}
   \Psi_{4A}
       \! &=& \!
      \bigl\{\bigl[[-1]|\!\Uparrow, \Uparrow, \Uparrow,\Uparrow \rangle
              +|\!\Downarrow,\Downarrow,\Downarrow,\Downarrow\rangle\bigr] 
                    \otimes |0\rangle
      \\ \nonumber &&
      +(\pm i)\bigl[|\!\Uparrow,\Uparrow,\Uparrow,\Downarrow\rangle+\dots\bigr]
                    \otimes|3\rangle
      \\ \nonumber &&
      +[-1](-1)\bigl[|\!\Uparrow,\Uparrow,\Downarrow,\Downarrow\rangle+\dots\bigr]
                    \otimes|2\rangle
      \\ \nonumber &&
      +(\mp i)\bigl[|\!\Uparrow,\Downarrow,\Downarrow,\Downarrow\rangle+\dots\bigr]
                    \otimes|1\rangle\bigr\}/4
\end{eqnarray}
and proper projection provides us with specifc entangled states with all
pseudo-spins aligned (Greenberger-Horne-Zeilinger states \cite{GHZ}), or
specific superpositions with exactly one-, two-, and three pseudospins
pointing downward, among them the Dicke states \cite{Dicke} with an equal
number of pseudospins pointing upward and downward. The generalization to
other values of $K$ is straightforward, including also cases with $n> 2^K$
producing a reduced but finite entanglement.

Letting the particles propagate beyond the line A, the many-particle
wavefunction undergoes mixing in the second beamsplitter, cf.\ Fig.\
\ref{fig:interferometer}. By properly choosing the transmission ($t =
\cos\theta$) and reflection ($r=i\sin\theta$) coefficients of the second
splitter, the pseudospins can be rotated into any direction, though all of the
pseudo-spins are rotated equally. Different rotations of the pseudo-spins can
be implemented by changing the characteristics of the splitter in time---the
time separation of the particle wavepackets can be enlarged, while
compromising between leaving sufficient time for the manipulation of the
splitter and keeping the system coherent.

For two particles, the Bell test for the pseudo-spin `singlet' state
$|\!\Uparrow,\Downarrow \rangle -|\!\Downarrow, \Uparrow \rangle$ is
particularly simple, as the four polarization angles $\theta_{1,2,1',2'}$ can
all be chosen to reside in the first quadrant, thus keeping the manipulation
of the beam splitter simple: the maximum violation is obtained for the usual
\cite{Aspect} angles $\theta_{12} = \theta_1-\theta_2 = \theta_{12'} =
\theta_{1'2} = \pi/8$ and $\theta_{1'2'} = 3\pi/8$ (the indices 1(2) refer to
the first (second) particle). The analysis of the pseudo-spin triplet states
involves angles in the second quadrant as well (as we have to replace
$\theta_2 \to -\theta_2$) and their experimental analysis is more demanding.

The action of the spin counter entangling two particles in a pseudo-spin
`singlet' state $|\!\Uparrow,\Downarrow \rangle -|\!\Downarrow, \Uparrow
\rangle$ can be observed in a simple experiment involving a finite
Aharonov-Bohm flux $\Phi$ through the interferometer. We inject the two (time
delayed, to avoid exchange effects) particles through the two different leads
on the left and measure the cross-correlator $\langle N_u N_d\rangle$ on the
right ($N_{u,d} \in \{0,1,2\}$ denote the number of particles observed in the
leads $u$ and $d$).  Without the counter, the product state generates the
result $\langle N_u N_d\rangle = P_{1u,1} P_{1d,2} +P_{1d,1}P_{1u,2}$, where
$P_{1x,i}$ denotes the probability to find the particle $i$ in the lead $x$.
With $P_{1u,1} = |t^2 + r^2 \exp(2\pi i\Phi /\Phi_0)|^2$ and $P_{1u,2} =
|r\,t[1+\exp(2\pi i\Phi /\Phi_0)]|^2$ and assuming symmetric splitters with
$t^2 = 1/2$, $r^2 = -1/2$, we find that $\langle N_u N_d\rangle =
[1+\cos^2(2\pi\Phi/\Phi_0)]/2$. On the other hand, with the spin-counter
selecting the singlet state, only paths where the combined trajectories
encircle the loop survive and the correlator is independent of $\Phi$,
$\langle N_u N_d\rangle = P_{2ud} = |t^4 + r^4|^2 + |2 t^2r^2|^2 = 1/2$ for
symmetric splitters. Hence post-selecting the spin-flipped events entangles
the particles and quenches the Aharonov-Bohm oscillations in the cross
correlator $\langle N_u N_d\rangle$. Such an analysis, although not as
rigorous as the classic Bell-inequality test but much simpler to implement,
may nevertheless serve as a preliminary indicator for the presence of
entanglement.

\section{Conclusion}

In conclusion, the simple spin counter in the `Gedanken Experiment' of full
counting statistics proves itself a fruitful idea---not only can it be
implemented as a real counting device with the help of quantum bits, its
generalization to many qubits combined with a non-trivial measurement
protocoll allows for the fabrication of a `quantum cardinality counter', where
a `primitive' physical information (rather than a binary one) can be
transformed into a binary form or directly used in the control and
manipulation of other---in particular quantum---information within a
mesoscopic setting.  On a more general level, we propose that restricting
ambitions to the design and implementation of special purpose devices may open
up new directions in quantum information processing which appear to be much
simpler to realize than the universal quantum computer, while still allowing
for interesting applications.

We thank Fabian Hassler, Andrey Lebedev, Renato Renner, Alexandre Blais, Denis
Vion, John Martinis, Patrice Bertet, and Fabien Portier for discussions and
acknowledge financial support by the CTS-ETHZ and the Russian Foundation for
Basic Research under grant No.\ 08-02-00767-a.

\end{document}